\documentclass[runningheads]{llncs}
\usepackage[T1]{fontenc}

\usepackage{graphicx}
\usepackage{url}
\usepackage{booktabs}
\usepackage{placeins}

\makeatletter
\renewcommand\section{\@startsection{section}{1}{\z@}%
  {-12pt plus -3pt minus -2pt}%
  {6pt plus 1pt minus 1pt}%
  {\normalfont\large\bfseries}}

\renewcommand\subsection{\@startsection{subsection}{2}{\z@}%
  {-10pt plus -2pt minus -1pt}%
  {4pt plus 1pt minus 1pt}%
  {\normalfont\bfseries}}
\makeatother

\begin{document}
\title{How Accurately Can the Energy Use of\\ Spark Applications Be Estimated Based\\on Resource Utilisation?}
\titlerunning{How Accurately Can the Energy Use of Spark Applications Be Estimated?}
\author{Youssef Moawad\inst{1} \and
Kathleen West\inst{1} \and
Vasilis Bountris\inst{2} \and
Philipp Thamm\inst{2} \and
\\Yehia Elkhatib\inst{1} \and
Lauritz Thamsen\inst{1}}
\authorrunning{Y. Moawad et al.}
\institute{
University of Glasgow, United Kingdom\\
\email{\{youssef.moawad, yehia.elkhatib, lauritz.thamsen\}@glasgow.ac.uk}\\
\email{k.west.1@research.gla.ac.uk}
\and
Humboldt-Universität zu Berlin, Germany\\
\email{\{vasilis.bountris, thammphx\}@hu-berlin.de}
}
\maketitle
\begin{abstract}

Distributed batch data processing applications are widely executed on cloud-based resources where restricted user access to node-level hardware energy counters hinders transparent sustainability accounting. Energy and carbon attribution methodologies therefore depend on power models and available resource utilisation traces, yet the accuracy of these estimates has to be validated while direct counters are available. In this work, we use Apache Spark running on Kubernetes as a case-study dataflow runtime and cluster resource manager to compare model-based energy estimates to Intel RAPL package and DRAM energy on an AWS bare-metal cloud and an on-premises cluster, comparing different CPU usage signals and memory coefficients. We show that external monitoring improves signed package-energy error relative to Spark task traces, reducing underestimation from -29.58\% to -24.41\% on AWS and from -24.00\% to -16.22\% on-premises.

\keywords{Energy Estimation, Power Models, Distributed Dataflows, Cluster Computing, Batch Data Processing, Sustainable Computing}
\end{abstract}

\section{Introduction}

Batch processing applications are widely used for data science, analytics, and machine learning workloads at scale~\cite{zaharia2016spark}, but their energy use is difficult to quantify once they run in virtualised environments~\cite{colmant2015processPower,smejkal2017eteam}. On bare-metal systems, hardware counters such as Intel's Running Average Power Limit (RAPL) can expose CPU and DRAM energy~\cite{khan2018rapl}. In cloud settings, energy and carbon footprint methodologies such as Cloud Carbon Footprint and Green Algorithms\footnote{\url{https://www.cloudcarbonfootprint.org/}, \url{https://www.green-algorithms.org/}} instead rely on power models and resource traces. Therefore, the central question that we address in this paper is: \textit{How accurate are energy use estimates based on linear power models and resource utilisation traces or node monitoring logs for distributed dataflow applications in cluster environments?}

We use RAPL to verify the power model methodology for this class of applications, using Apache Spark~\cite{zaharia2016spark} applications as a case study. Spark event logs expose task-level metrics~\footnote{\url{https://spark.apache.org/docs/latest/monitoring}, accessed: 2026-05-12}, which can be converted into task traces and which we pass to Ichnos, a trace-based energy and carbon estimator for cluster computing applications~\cite{west2025ichnos,west2026ichnosplus}. Seeing cloud systems can often access process-level CPU metrics from \texttt{/proc}, without elevated permissions, task-level utilisation could stem either from application event logs or external process monitoring.
We, therefore, compare Spark task trace-derived energy estimates with those derived from node-level process monitoring.  We also consider the energy from memory utilisation and its effect on the accuracy of the total estimated energy for the workload.

Validating the accuracy of utilisation-derived energy estimates for distributed data processing is important before the estimates are used further for carbon accounting, program optimisation, scheduling, or resource allocation decisions.

We make the following contributions in this paper:
\begin{itemize}
    \item We formulate a methodology to construct comparable energy use estimates for distributed dataflow applications through two different CPU usage signals, validating node-specific linear power models directly against hardware-level RAPL readings as a baseline.
    \item We empirically evaluate this methodology across 13 BenchSpark~\cite{will2024benchsparkAutoscaling} applications deployed on two contrasting Kubernetes infrastructures: an AWS bare-metal cluster and an on-premises commodity cluster.
    \item We compare different power model calibrations for generating the energy estimates in two settings: one calibrated under idle system conditions, and another under minimal infrastructure readiness.
    \item We discuss why estimates should be interpreted component-wise instead of as aggregates, as CPU and DRAM errors can cancel out in total energy.
\end{itemize}

\section{Background}

This section describes the runtime, carbon-accounting, and power-model concepts relevant for our study.

\subsection{Spark Application Model}

Distributed batch processing systems execute bounded workloads across multiple machines. Distributed dataflow systems, such as Spark or Apache Flink, and Apache Beam\footnote{\url{https://flink.apache.org/}, \url{https://beam.apache.org/}}, are one way to implement this model: input data is partitioned, transformations are applied to partitions in parallel, and work is scheduled as data-parallel tasks. Spark applications consist of a driver process that manages executors which run tasks across the worker nodes. Spark jobs are decomposed into stages, which are themselves made up of tasks, the most fine-grained unit of work in Spark which can be scheduled on an executor. Tasks in a single stage may be scheduled to run across different executors, across different worker nodes; but a single task will run on one executor and will by default be capped to one CPU core.

\subsection{Cloud Carbon Footprint and Green Algorithms}

The Cloud Carbon Footprint (CCF) methodology estimates energy consumption based on cloud resource utilisation, linear processor power models, memory coefficients, and grid carbon intensity. Green Algorithms offers a more general computational carbon emissions estimator, relying primarily on user-supplied hardware specifications or thermal design power assumptions of processor manufacturers. Both follow the same broad resource usage to energy to carbon flow.

\section{Methodology}

We align the RAPL baseline, trace signals, and measured node models into comparable package, DRAM, and total-energy estimates. For this, RAPL counters are collected independently of Spark, while an external monitor gathers process snapshots, during each workload run.

\subsection{Establishing Node-Specific Power Models}
\label{sec:measuring_power_models}

On each node, we measure the CPU power model using \texttt{turbostress} to generate controlled CPU loads from $0\%$ to $100\%$ in $10\%$ increments. At each point, \texttt{turbostat} reads RAPL counters and derives package wattage from the energy delta over the measurement interval. We average three stress runs per load point and fit the averaged points to a linear model.

\begin{figure}
\centering
\includegraphics[width=0.97\linewidth]{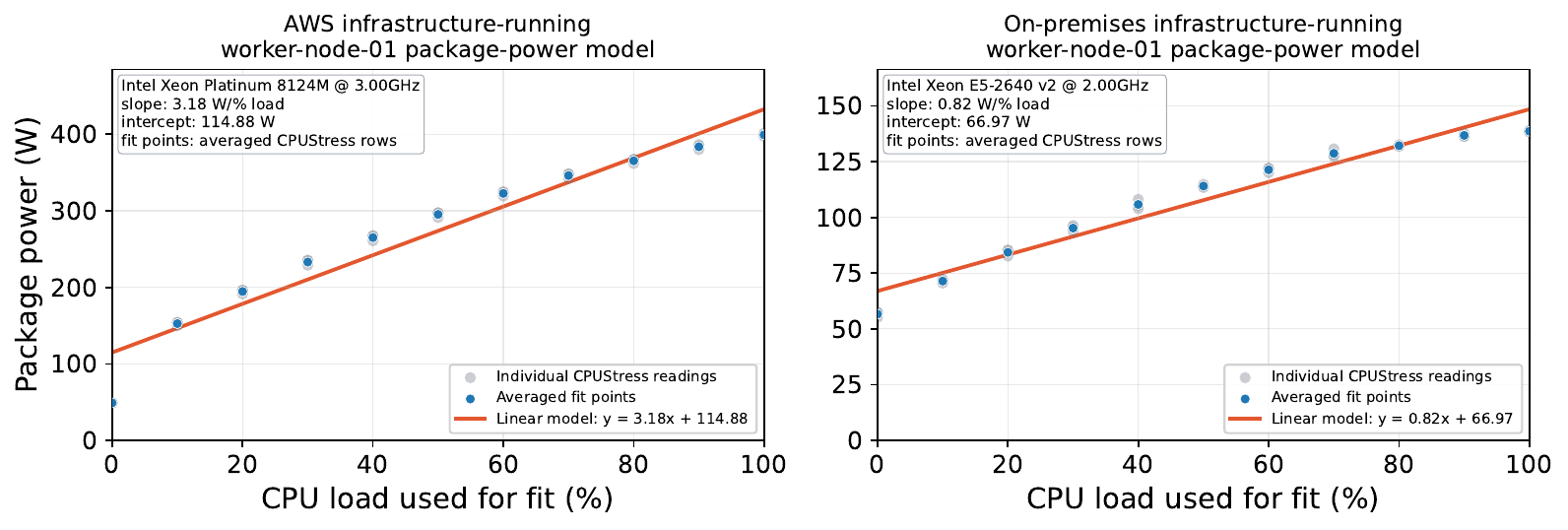}
\caption{Representative package-power model fits. Each plot contrasts RAPL package-power samples at controlled CPU-load points with the fitted per-node linear model used for energy estimation.}
\label{fig:power-model-samples}
\end{figure}

Memory coefficients are measured separately by stressing memory at the same $10\%$ increments and reading DRAM energy from RAPL. The coefficient is reported as \texttt{W/GB} by dividing observed DRAM watts by total node memory capacity, matching the coefficient form used in allocation-based accounting methods like CCF. We average the loaded stress points rather than using idle memory power or a single point because DRAM power has a smaller dynamic range than package power once memory is active. Figure~\ref{fig:power-model-samples} shows representative fitted package-power models for the two evaluated environments.

For each node, we calibrate one model with Kubernetes infrastructure services running and another with said services stopped. The main results use the former calibration state, with the latter being used as a sensitivity check.

\subsection{Ichnos Trace Format and Energy Estimation}
\label{sec:IchnosEnergyEstimation}

For task-level energy consumption estimation, we use Ichnos, which was developed to estimate the energy consumption and carbon emissions of Nextflow scientific workflows running on compute clusters~\cite{diTommaso2017nextflow,west2026systematicCarbonAware}.
Ichnos utilises linear power models, derived as described in Section~\ref{sec:measuring_power_models}, to estimate energy consumption using CPU usage, and in turn carbon emissions, using carbon intensity data. It consumes task or process-based traces and applies the node-specific power models over each trace interval.

The Ichnos trace format is a tabular file whose rows contain a host name, start and end time, CPU usage, and memory usage. Ichnos then applies node-specific power models to convert CPU usage into estimated package power during that task runtime, and can compute energy by integrating power over the task interval. Memory energy usage is computed separately using either the measured memory coefficient or the default CCF value.

\subsection{Spark Task CPU Trace}

We convert Spark executor task events into Ichnos trace rows using task launch time, finish time, and executor CPU time. The CPU value is computed as:
\[
\begin{array}{rcl}
\mathrm{duration}_{ms} &=& \mathrm{finish}_{ms}-\mathrm{launch}_{ms},\\
\mathrm{trace\_cpu} &=& (\mathrm{CPUTime}_{ns}/10^6) / \mathrm{duration}_{ms} \times 100.
\end{array}
\]
This is a percentage value out of one core, not out of the entire CPU, which is what Ichnos expects.

\subsection{External Monitoring Trace}

We use an external monitor that polls \texttt{/proc} every 5 seconds and records process snapshots. Specifically, for each process, the monitor records command and cgroup identifiers together with cumulative CPU time, and Spark executor processes are selected by matching executor-related command and cgroup markers. Between consecutive snapshots, the monitor computes process CPU-time deltas and divides them by the wall-clock interval, producing fixed-window CPU trace rows. This signal differs from Spark event-log traces: it is sampled at fixed wall-clock intervals and can include executor activity that is not attributed to completed Spark tasks, but it can also miss short-lived tasks, making this sensitive to the chosen polling interval. We choose 5 seconds as a compromise between monitoring overhead and temporal resolution. While short executor bursts may be missed, this interval provides repeated observations for most executors.

\subsection{RAPL and Monitoring Power Consumption}

We collect RAPL energy by reading the \texttt{package} and \texttt{dram} domains once per second from before job submission until completion. Multiple sockets are logged and later summed, and package and DRAM readings are stored separately.

During analysis, the RAPL logs are unwrapped and normalised into cumulative energy series. For each trace interval, we interpolate the cumulative value at the start and end, subtract the two, convert to kWh, and sum across intervals and nodes. Spark task traces are compared with RAPL over task intervals, while external monitoring traces are compared over the 5-second monitoring windows.

We validate only worker-node package and DRAM energy over trace-aligned intervals because the model estimates workload energy from Spark task and process traces on worker nodes. Control-plane energy is outside this boundary.

For computing memory energy, we use total node memory as the allocation amount because the experimental clusters are dedicated to these runs. We then evaluate two memory coefficients using this value: the measured node's memory coefficient (as described in Section~\ref{sec:measuring_power_models}), and the default CCF memory coefficient.

\section{Experimental Setup}

We evaluate the methodology on two dedicated clusters: an AWS bare-metal cluster with four EC2 \texttt{c5n.metal} nodes, and an on-premises Kubernetes cluster using eight Spark worker nodes. Both clusters run Ubuntu 22.04 LTS, Apache Spark 3.5.5, and Kubernetes 1.32, with storage provided by Rook-managed CephFS using Ceph 19.2. These environments differ in processor generation, memory capacity, and executor layout, giving two distinct validation settings. The AWS cluster uses one control node and three Spark workers; each worker exposes an Intel Xeon Platinum 8124M CPU at 3.00GHz, 72 hardware threads, and 192GB memory, with Spark configured for three executors using 64 cores and 120GB memory each. The on-premises cluster uses \texttt{worker-node-01}--\texttt{08} as Spark workers; each worker has an Intel Xeon E5-2640 v2 CPU at 2.00GHz, 32 hardware threads, and 64GB memory, with Spark configured for eight executors using 30 cores and 32g memory each.

For our target workloads, we use the BenchSpark suite~\cite{will2024benchsparkAutoscaling}. We evaluate search, sorting, shuffle-heavy, text-processing, and iterative machine-learning jobs. The AWS runs use larger input configurations than the on-premises runs because the AWS nodes have substantially higher capacity. Table~\ref{tab:workload-scale} reports the workload input configurations and median runtimes; all rows in the result tables are medians over three completed runs.

\begin{table}[!htbp]
\caption{Workload configurations and median runtimes over three runs; slash-separated values denote AWS/on-premises.}
\label{tab:workload-scale}
\centering
\scriptsize
\setlength{\tabcolsep}{3pt}
\renewcommand{\arraystretch}{0.9}
\begin{tabular}{@{}p{0.23\linewidth}rrp{0.48\linewidth}@{}}
\toprule
& \multicolumn{2}{c}{Runtime (s)}
& Input configuration \\
\cmidrule(lr){2-3}
Workload & AWS & On-prem. & (AWS / on-premises) \\
\midrule
\texttt{grep-medium}
& 679.30 & 238.14
& 3.0B / 2.0B rows, 80 B/record, match 50k \\

\texttt{groupby-medium}
& 23.27 & 40.56
& 120M / 80M records \\

\texttt{join-medium}
& 355.35 & 115.96
& 150M / 60M visits, 900M / 360M pages \\

\texttt{kmeans-large}
& 7383.87 & 2864.39
& 200M / 125M points, 100 dimensions, 100 clusters, 100 iterations \\

\texttt{kmeans-medium}
& 3796.64 & 1111.91
& 150M / 78.125M points, 100 dimensions, 100 clusters, 100 iterations \\

\texttt{kmeans-small}
& 438.98 & 24.63
& 25M / 10k points, 100 dimensions, 10 clusters, 100 iterations \\

\texttt{linreg-medium}
& 948.57 & 129.08
& 240M / 60M points, 50 features, 50 iterations \\

\texttt{logreg-medium}
& 1245.13 & 3829.17
& 240M / 60M points, 50 features \\

\texttt{select-medium}
& 4723.98 & 275.76
& 18.0B / 1.8B rows \\

\texttt{sort-medium}
& 2021.22 & 75.53
& 1.6B / 160M rows, 100 partitions \\

\texttt{sort-large}
& 3770.84 & 139.62
& 2.4B / 320M rows, 100 partitions \\

\texttt{sort-small}
& 681.26 & 27.85
& 850M / 85M rows, 100 partitions \\

\texttt{wordcount-medium}
& 24.86 & 14.46
& 1.5B / 300M records \\
\bottomrule
\end{tabular}
\end{table}

\section{Results}

We present the primary results derived using models calibrated with infrastructure services running, and compare package, DRAM, and package-plus-DRAM total error against their RAPL baseline target. All reported values are signed percentage errors over workload-level medians, with standard deviations calculated across workloads. Workload-level plots show median signed errors, with error bars spanning the minimum and maximum observed values across completed runs. Each estimate is shown against RAPL energy integrated over the corresponding trace windows. Negative signed errors denote underprediction relative to the relevant RAPL target. Absolute energy values for all workloads are available in the accompanying artifact~\cite{pecs2026artifact}.

\subsection{CPU Package Results}

Table~\ref{tab:cpu-signal-aggregate} summarises package-energy error across the selected workloads. We see that external monitoring reduces mean package-energy underprediction in both environments compared with Spark task traces.

\begin{table}
\caption{Package-energy signed percentage error relative to RAPL package energy by CPU signal for models calibrated with infrastructure services running.}
\label{tab:cpu-signal-aggregate}
\begin{minipage}{0.88\linewidth}
\centering
\scriptsize
\setlength{\tabcolsep}{3pt}
\centering
\begin{tabular}{@{}p{0.46\linewidth}rr@{}}
\toprule
Environment & \shortstack{Spark package (\%)} & \shortstack{External package (\%)} \\
\midrule
AWS bare-metal & \(-29.58 \pm 16.12\) & \(-24.41 \pm 10.56\) \\
On-premises & \(-24.00 \pm 11.47\) & \(-16.22 \pm 12.13\) \\
\bottomrule
\end{tabular}
\end{minipage}
\end{table}

\begin{figure}
\centering
\begin{minipage}{0.92\linewidth}
\includegraphics[width=\linewidth]{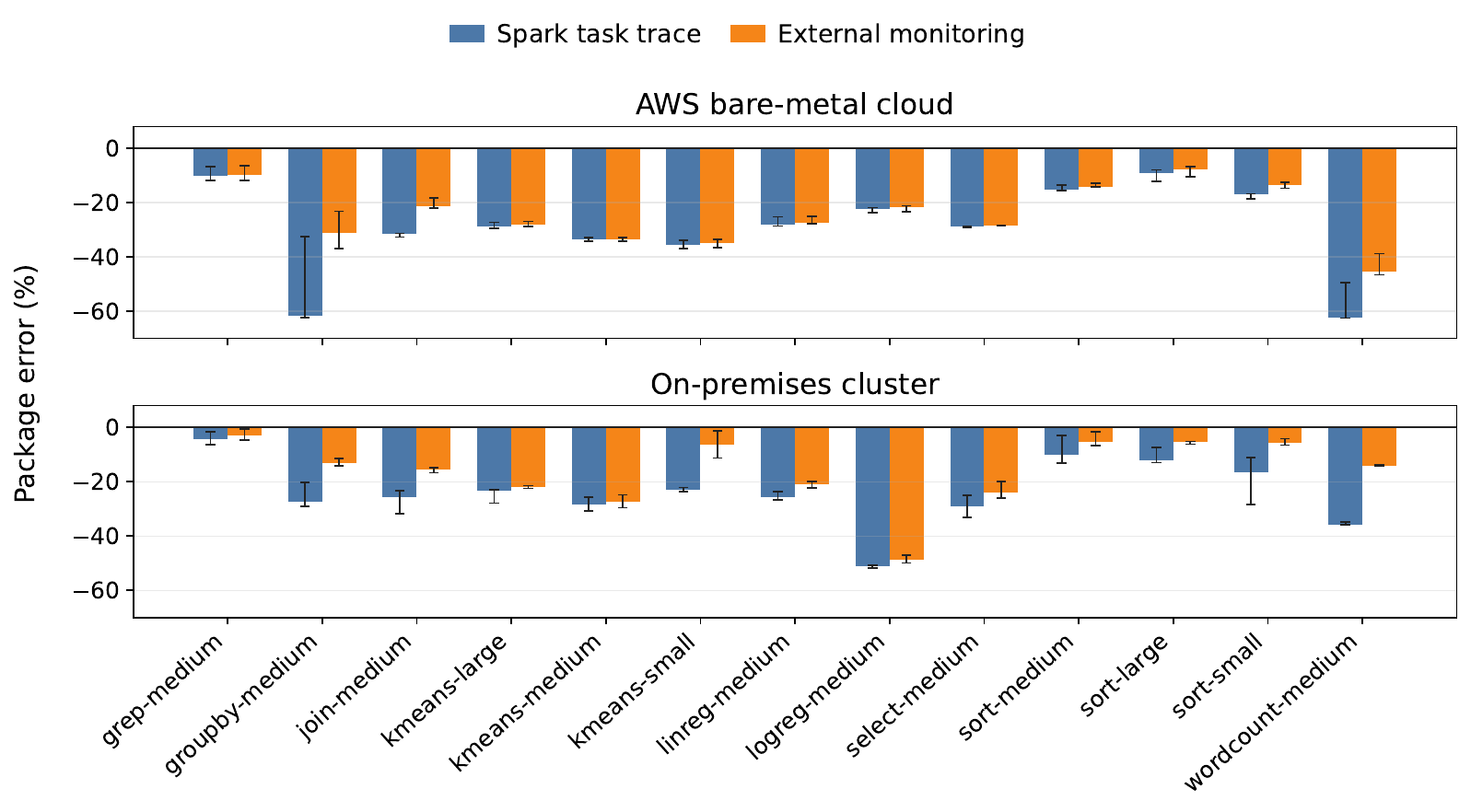}
\end{minipage}
\caption{Workload-level package-energy error relative to RAPL, comparing CPU usage across Spark task trace and external monitoring for models calibrated with infrastructure services running.}
\label{fig:package-workload}
\end{figure}

Figure~\ref{fig:package-workload} shows the workload-level pattern behind the aggregate.
Across this range, the external-monitoring estimates generally lie closer to their matched RAPL values than the Spark task-trace estimates, as reported through the normalised errors in Table~\ref{tab:cpu-signal-aggregate}. The absolute values show why short workloads can have large percentage errors despite comparatively small energy differences.
Longer workloads do not all follow the same pattern: \texttt{kmeans-large} and \texttt{select-medium} remain around -28\% under both signals, while \texttt{sort-large} reaches -7.71\% with external monitoring. Runtime therefore helps explain some fixed-cost effects, but it is not the only determinant of package-energy error.

\subsection{Memory Results}

Table~\ref{tab:memory-aggregate} compares memory energy errors against RAPL DRAM. On AWS, the default CCF coefficient substantially reduces DRAM underprediction. On-premises, both overpredict DRAM, although the default produces a smaller error.

\begin{table}
\caption{Memory-coefficient error for Total Node Memory against RAPL DRAM where the measured coefficient is calibrated with infrastructure services running.}
\label{tab:memory-aggregate}
\centering
\scriptsize
\setlength{\tabcolsep}{4pt}
\begin{minipage}{0.88\linewidth}
\centering
\begin{tabular}{@{}p{0.45\linewidth}rr@{}}
\toprule
Environment & Measured coeff. (\%) & Default CCF coeff. (\%) \\
\midrule
AWS bare-metal cloud & \(-17.86 \pm 7.34\) & \(-1.54 \pm 8.80\) \\
On-premises cluster & \(+77.78 \pm 27.80\) & \(+54.73 \pm 24.18\) \\
\bottomrule
\end{tabular}
\end{minipage}
\end{table}

\begin{figure}
\centering
\begin{minipage}{0.92\linewidth}
\centering
\includegraphics[width=\linewidth]{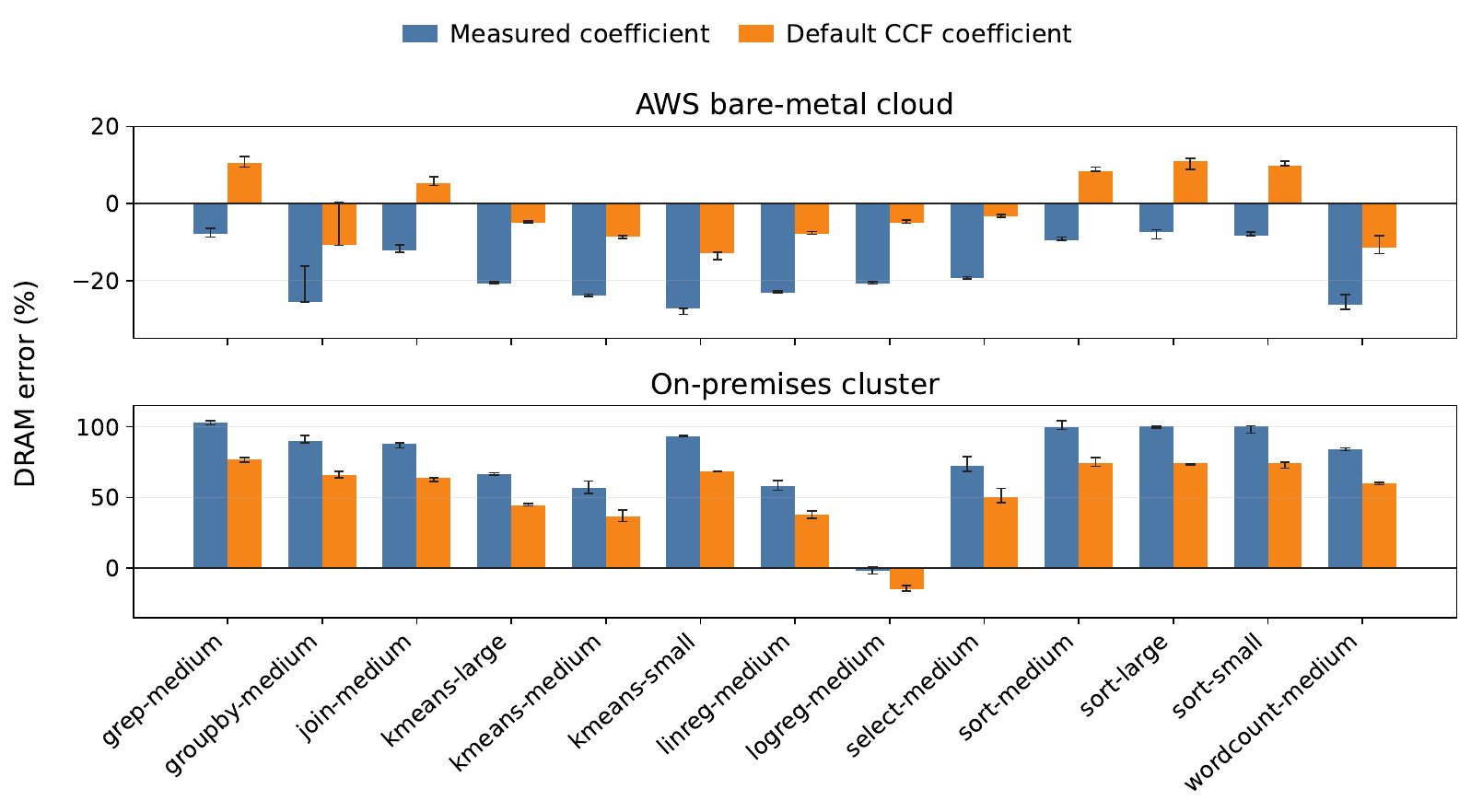}
\end{minipage}
\caption{Workload-level DRAM relative to RAPL DRAM error for Total Node Memory with measured memory coefficients and with the default CCF cloud memory coefficient.}
\label{fig:memory-workload}
\end{figure}

Figure~\ref{fig:memory-workload} shows the memory energy error values for the different workloads. AWS memory errors are mostly negative under the measured coefficient and centered much closer to zero under the default coefficient. For on-premises, memory errors are strongly positive for most workloads under both coefficients, with \texttt{logreg-medium} as the main exception.

\subsection{Total Energy}

Table~\ref{tab:total-aggregate} reports package-plus-DRAM error for the four CPU-signal and memory-coefficient combinations. On AWS, default CCF memory improves total-energy error for both CPU signals. For on-premises, the default coefficient makes the total more negative because the measured-memory total had benefited from cancellation between negative package error and positive DRAM error.

\begin{table}[!htbp]
\caption{Package-plus-DRAM signed percentage error relative to RAPL total energy using models calibrated with infrastructure services running. Values are mean \(\pm\) standard deviation.}
\label{tab:total-aggregate}
\centering
\scriptsize
\setlength{\tabcolsep}{2pt}
\renewcommand{\arraystretch}{0.98}
\begin{tabular}{@{}p{0.27\linewidth}cccc@{}}
\toprule
Environment & \multicolumn{2}{c}{Spark trace} & \multicolumn{2}{c}{External monitoring} \\
\cmidrule(lr){2-3}\cmidrule(lr){4-5}
& \shortstack{Measured\\(\%)} & \shortstack{Default\\(\%)}
& \shortstack{Measured\\(\%)} & \shortstack{Default\\(\%)} \\
\midrule
AWS bare-metal cloud
& \(-31.61 \pm 13.61\)
& \(-26.49 \pm 14.29\)
& \(-22.43 \pm 9.62\)
& \(-17.20 \pm 10.33\) \\
On-premises cluster
& \(-25.34 \pm 13.11\)
& \(-29.50 \pm 12.97\)
& \(-0.95 \pm 14.06\)
& \(-4.71 \pm 13.59\) \\
\bottomrule
\end{tabular}
\end{table}

\begin{figure}[!t]
\centering
\begin{minipage}{0.92\linewidth}
\centering
\includegraphics[width=\linewidth]{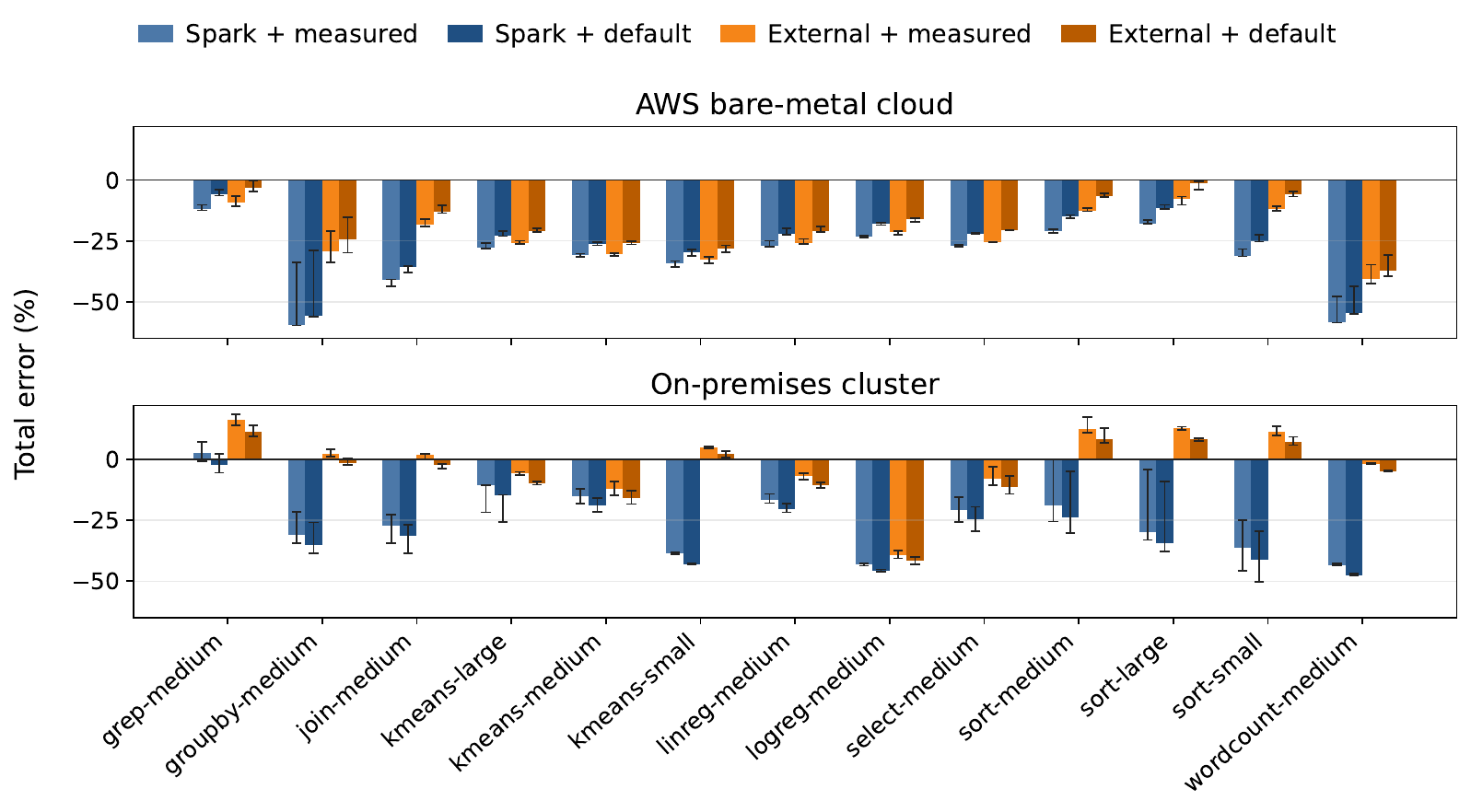}
\end{minipage}
\caption{Workload-level total energy error relative to RAPL package-plus-DRAM total energy for the four CPU-signal and memory-coefficient combinations using models calibrated with infrastructure services running.}
\label{fig:total-workload}
\end{figure}

Figure~\ref{fig:total-workload} shows why total energy should be interpreted after inspecting package and DRAM separately: total error can improve because one component becomes more accurate, or because two component errors cancel.

\subsection{Sensitivity to Power Model Calibration}

The main results use power models calibrated with Kubernetes infrastructure services running because the workloads execute under those conditions. To test sensitivity to this choice, Table~\ref{tab:calibration-sensitivity} reports package-energy errors when the same CPU signals are evaluated with models calibrated with those services stopped. These models make package errors more negative in both environments and for both CPU signals, so the model calibrated with services running is the more appropriate choice for these experiments.

\begin{table}[!htbp]
\caption{Package-energy sensitivity to calibration state, reported as signed percentage error relative to RAPL package energy.}
\label{tab:calibration-sensitivity}
\centering
\scriptsize
\setlength{\tabcolsep}{4pt}
\begin{tabular}{@{}p{0.3\linewidth}rrrr@{}}
\toprule
Environment & \multicolumn{2}{c}{Spark trace} & \multicolumn{2}{c}{External monitoring} \\
\cmidrule(lr){2-3}\cmidrule(lr){4-5}
& Running (\%) & Stopped (\%) & Running (\%) & Stopped (\%) \\
\midrule
AWS bare-metal cloud & -29.58 & -39.55 & -24.41 & -35.21 \\
On-premises cluster & -24.00 & -27.94 & -16.22 & -20.57 \\
\bottomrule
\end{tabular}
\end{table}

\FloatBarrier

\section{Discussion}

This section interprets the results and summarises the main threats to validity.

\subsection{Discussion of Accuracy}

This paper primarily asks how accurately linear power models estimate the energy use of distributed dataflow workloads. We tested different analysis methods and found that accuracy depends on factors including the utilised CPU usage signal, the chosen memory coefficient, and the state of the cluster when the power models were calibrated.

For CPU energy, external process monitoring improves package estimates in both environments, from -29.58\% to -24.41\% on AWS and from -24.00\% to -16.22\% for on-premises. This suggests that process-level monitoring captures more executor activity than the application's task-attributed CPU time. Both signals still underestimate the measured CPU package energy on average; which is expected because neither captures all static CPU power, full platform infrastructure, monitoring overhead, or other node activity during workload execution.

Relatedly, we do not claim that CPU and memory traces alone can perfectly reconstruct the complete energy consumption. Still, validating this restricted model is worthwhile as it is often the first step of carbon footprint methodologies; power usage effectiveness and carbon intensity can then be applied subsequently.

Memory behaves differently across the two environments, which strongly affects total energy estimates. AWS DRAM is underestimated with the measured coefficient and nearly centered with the default coefficient, while on-premises DRAM is substantially overestimated with both coefficients. The on-premises near-zero total error is therefore partly due to cancellation between positive DRAM error and negative package error; on AWS, total error remains negative.

The calibration sensitivity results in Table~\ref{tab:calibration-sensitivity} show that the initial power model calibration conditions affect the estimated energy. Since the workloads run with Kubernetes infrastructure services active, the model calibrated with those services running is the more appropriate choice for these experiments.

\subsection{Threats to Validity}

The evaluation uses only selected BenchSpark applications.
These workloads cover search, sorting, shuffle-heavy jobs, text processing, and iterative machine learning, but they do not cover streaming, interactive analytics, accelerated workloads, or multi-tenant deployments. Spark is also the only dataflow runtime evaluated; other systems such as Flink or Beam may expose different logs, scheduling behaviour, and process structure.

The infrastructure scope is also limited to two environments: one AWS bare-metal cloud and one on-premises cluster. This is enough to expose environment-dependent behaviour, especially for DRAM, but not enough to claim generality across processor generations, memory systems, cloud instance families, or governor policies. The validation baseline is limited to RAPL package and DRAM domains rather than wall-plug energy, and we have not accounted for control-plane energy use. The CPU trace signals have different granularities and semantics: event-log traces are task-attributed, while process traces are sampled at fixed intervals and can miss short-lived spikes. The external monitor samples every 5 seconds, so short executor bursts may be smoothed or missed, particularly for short tasks. We report workload-level medians and ranges rather than formal significance tests because the study has small per-workload repetition counts and heterogeneous workload families. The linear power models are interpretable and match common cloud-accounting methodology, but they do not fully capture frequency scaling, thermal state, or nonlinear package power behaviour.

Despite these threats, our experiments show that carefully configured linear power models can be used for meaningful energy consumption estimates.

\section{Related Work}

Our work intersects three topics: energy use of distributed dataflow systems, hardware counter-based energy estimation, and trace-driven power modelling.

Case studies of Spark and MapReduce have examined energy-performance trade-offs, forecasting, resource management, and scheduling for big data workloads~\cite{mashayekhy2015mapreduceEnergy,maroulis2017sparkEnergyScheduling,li2019sparkEnergyScheduling,volpini2026sparkEnergy,gonzalezOrdiano2018sparkEnergyForecasting,aziz2019sparkResourceManagement}.
We instead focus on validating trace-derived energy estimates against RAPL to provide evidence supporting using such estimates in settings without RAPL access.

RAPL is widely used for processor and DRAM energy measurement, but it requires careful interpretation of domains, sampling, and counter handling~\cite{khan2018rapl}.
EnergAt attributes CPU and DRAM energy at thread level in multi-tenant systems~\cite{he2024energat}, while Nf-PEAK maps Kubernetes processes to Nextflow tasks and attributes RAPL-based energy in shared cluster environments~\cite{thamm2026nfpeak}. Power models offer an alternative to energy measurements, but their use requires careful validation against measurements~\cite{davis2012chaos,colmant2018next700,fahad2019comparativeEnergy}. We therefore use RAPL in this paper to validate estimates derived from Spark and process traces, targeting cloud environments where node-level RAPL is normally unavailable.

Trace-driven and model-based power meters such as Joulemeter and PowerAPI estimate energy from resource usage where direct per-application measurement is not available~\cite{kansal2010joulemeter,colmant2018next700,fieni2024powerapi}. Ichnos similarly uses node-specific fitted power models for utilisation-based estimates. Ichnos also provides the trace format and trace-to-energy implementation used here~\cite{west2025ichnos}. We extend this line of work by comparing the use of Spark event-log traces and external process-monitoring traces and by validating package and DRAM estimates component-wise across two infrastructures.

\section{Conclusion}

We set off to empirically explore the accuracy of node-specific linear power models for estimating the energy use of distributed batch data processing workloads, using Spark as a case-study dataflow runtime. We showed that the estimation accuracy depends on model calibration state, CPU signal, and memory-coefficient assumption as well as the workload. For CPU energy, external process monitoring reduces mean signed package-energy error to -24.41\% on an AWS bare-metal cluster and -16.22\% on the on-premises cluster. Memory estimation is more environment-sensitive: the default CCF memory coefficient improves AWS total-energy error, but worsens the on-premises total because it reduces cancellation between negative package error and positive DRAM error. We conclude, therefore, that CPU usage data is useful for estimating the energy of distributed batch processing on cloud-based clusters, but package, DRAM, and total energy should be interpreted component-wise before carbon-footprint accounting.

\begin{credits}

\paragraph{\ackname}
This work was supported by the Engineering and Physical
Sciences Research Council under grant number UKRI154
("Casper: Carbon-Aware Scalable Processing in Elastic
Clusters"). We further thank AWS for cloud credits.

\paragraph{\discintname}
We have no competing interests to declare.

\paragraph{Rights retention statement.}
For the purpose of open access, we have applied a Creative Commons Attribution (CC BY) license to any Author Accepted Manuscript version arising from this submission.

\paragraph{AI assistance disclosure.}
We used AI-based tools for language editing and assistance with LaTeX and code. We reviewed and verified all outputs and take full responsibility for the scientific claims, analyses, code, and final manuscript.

\paragraph{Artifact availability.}
We make the experimental execution scripts, raw data, and plotting code available in the accompanying artifact~\cite{pecs2026artifact}.

\end{credits}
\bibliographystyle{splncs04}
\bibliography{references}
\end{document}